\documentclass[12pt]{article}
\usepackage{amssymb,amsmath,epsfig}
\allowdisplaybreaks

\begin{document}

\title{\bf Non-minimal Coupling of Torsion-matter Satisfying Null Energy Condition for Wormhole Solutions}
\author{Abdul Jawad \thanks {jawadab181@yahoo.com, abduljawad@ciitlahore.edu.pk} and
Shamaila Rani \thanks {shamailatoor.math@yahoo.com, drshamailarani@ciitlahore.edu.pk}\\
Department of Mathematics, COMSATS Institute of\\ Information
Technology, Lahore-54000, Pakistan.}
\date{}

\maketitle
\begin{abstract}
We explore wormhole solutions in a non-minimal torsion-matter
coupled gravity by taking an explicit non-minimal coupling between
the matter Lagrangian density and an arbitrary function of torsion
scalar. This coupling depicts the transfer of energy and momentum
between matter and torsion scalar terms. The violation of null
energy condition occurred through effective energy-momentum tensor
incorporating the torsion-matter non-minimal coupling while normal
matter is responsible for supporting the respective wormhole
geometries. We consider energy density in the form of
non-monotonically decreasing function along with two types of
models. First model is analogous to curvature-matter coupling
scenario, that is, torsion scalar with $T$-matter coupling while the
second one involves a quadratic torsion term. In both cases, we
obtain wormhole solutions satisfying null energy condition. Also, we
find that the increasing value of coupling constant minimizes or
vanishes the violation of null energy condition through matter.
\end{abstract}

\section{Introduction}

The topological handles which connect distant regions of the
universe as a bridge or tunnel is named as wormhole. The most
amazing thing is the two-way travel through wormhole tunnel which
happened when throat remains open. That is, to prevent the wormhole
from collapse at non-zero minimum value of radial coordinate. In
order to keep the throat open, the exotic matter is used which
violate the null energy condition and elaborated the wormhole
trajectories as hypothetical paths. The violation of null energy
condition is the basic ingredient to integrate wormhole solutions.
This configuration is firstly studied by Flamm \cite{1} and then
leads Einstein and Rosen \cite{2} to contribute into the successive
steps for the construction of wormhole solutions. The work of Morris
and Thorne \cite{3} evoked the wormhole scenario and leads to the
new directions. The usual matter are considered to satisfy the
energy conditions, therefore, some exotic type matter is employed
for these solutions. There exists some wormhole solutions in
semi-classical gravity through quantum effects such as Hawking
evaporation and Casimir effects \cite{3,4} where energy conditions
are violated. One may take some such types of matter which acted as
exotic matter, for instance, phantom energy \cite{5}, tachyon matter
\cite{6}, generalized Chaplygin gas \cite{7}, some non-minimal
kinetic coupling, etc.

In order to find some realistic sources that support the wormhole
geometry or minimize the usage of exotic matter, different variety
of wormhole solutions are explored. This includes thin-shell,
dynamical and rotating wormholes \cite{8}. However, the
concentration goes towards modified theories of gravity where the
effective scenario gives the violation of null energy condition and
matter source supports the wormholes. Bronnikov and Starobinsky
\cite{8+} proved a general no-go theorem which yields no wormholes
(static or dynamic) can be constituted in any ghost-free
scalar-tensor theory of gravity (which also includes $f(R)$ gravity
as a particular case), under some conditions on the non-minimal
coupling function. In $f(R)$ gravity, Lobo and Oliveira \cite{9}
found that the higher curvature terms in effective energy-momentum
tensor are responsible for the necessary violation for the wormhole
solutions in ghost-containing scenario. They assumed a particular
shape function along with various fluids to check the validity of
energy conditions. Jamil et al. \cite{10} discussed several static
wormhole solutions in this gravity with a noncommutative geometry
background through Gaussian distribution. They considered power-law
solution as first and construct wormhole geometry as well as
analyzed the validity of energy conditions. Secondly, they explored
these solutions with the help of shape function. Taking into account
Lorentzian distribution of energy density, Rahaman et al. \cite{11}
derived some new exact solutions under same manner and gave some
viable wormhole solutions.

In extended teleparallel gravity being the modification of
teleparallel gravity \cite{ft,ft2}, static as well as dynamical
wormhole solutions are also explored. In this way, B\"{o}hmer et al.
\cite{12} investigated wormhole solutions in this gravity by taking
some specific $f(T)$ forms, shape function as well as redshift
functions which are the basic characteristics of these solutions.
Assuming different fluids such as baro-tropic, isotropic and
anisotropic, Jamil et al. \cite{13} enquired the possibility of some
realistic sources for wormhole solutions. Sharif and Rani explored
the wormhole solutions in this gravity taking noncommutative
background with Gaussian distribution \cite{14}, dynamical wormhole
solutions \cite{18}, for the traceless fluid \cite{15}, with the
inclusion of charge \cite{16} and galactic halo scenario \cite{17}.
They considered some power-law $f(T)$ functions for which effective
energy-momentum tensor depending on torsion contributed terms
violated the null energy condition. Recently, Jawad and Rani
\cite{19} constructed wormhole solutions via Lorentzian distribution
in noncommutative background. They concluded that their exist the
possibility of some realistic wormhole solutions satisfying energy
conditions and stayed in equilibrium. We also studied some higher
dimensional wormhole solutions in Einstein Gauss-Bonnet gravity
\cite{20}.

The modification of theories taking some non-minimal coupling
between matter and curvature becomes center of interest now-a-days.
There exist such theories involving these coupling as
$f(R)=f_1(R)+[1+\lambda f_2(R)]\mathcal{L}_m$ and
$f(R,\mathcal{L}_m)$, \cite{20+} etc. Harko et al. \cite{20++}
introduced the most general conditions in the framework of modified
gravity, that is the matter threading the wormhole throat satisfies
all of the energy conditions while the gravitational fluid (such as
higher order curvature terms) support the nonstandard wormhole
geometries. They explicitly showed that wormhole geometries can be
theoretically constructed without the presence of exotic matter but
are sustained in the context of modified gravity. Taking into
account some specific cases of modified theories of gravity, namely,
$f(R)$ gravity, the curvature-matter coupling and the
$f(R,\mathcal{L}_m)$ generalization, they showed explicitly that one
may choose the parameters of the theory such that the matter
threading the wormhole throat satisfies the energy conditions.

Following the same scenario as for $f(R)$ theory, Harko et al.
\cite{21} proposed non-minimal torsion-matter coupling as a
extension of $f(T)$ gravity. In this gravity, two arbitrary
functions $f_1(T)$ and $f_2(T)$ are introduced such as $f_1$ is the
extension of geometric part while $f_2$ is coupled with matter
Lagrangian part through some coupling constant. They discussed this
theory for cosmological aspects of evolving universe and deduced
that the universe may represent quintessence, phantom or crossing of
phantom-divide line, inflationary era, de Sitter accelerating phase,
in short, a unified description. In this gravity, Nashed \cite{22},
Feng et al. \cite{23} and Carloni et al. \cite{24} studied
spherically symmetric solutions, cosmological evolutions and
compared their results with observational data and phase space
analysis, respectively. Garcia and Lobo \cite{25} explored wormhole
solutions by non-minimal curvature-matter coupling taking linear
functions, $f_1=R,~f_2=R$. They concluded that the wormhole
solutions in realistic manner depends on higher values of coupling
parameter.

The paper has following symmetry. In next section 2, we describe the
$f(T)$ gravity and non-minimal torsion-matter coupling extension.
Section 3 is devoted to the gravitational field equations for
wormhole geometry in the coupling scenario. We find the general
conditions on matter part for the validity of null energy condition.
Also, we examined the effective energy-momentum tensor being the
responsible for violation of null energy condition. In section 4, we
explore wormhole solutions taking into account two well-known
models. These models involving linear torsion scalar coupled with
$T$-matter and quadratic torsion term with matter represented
wormhole geometry. In the last section, we summarize the results.

\section{Non-minimal Torsion-matter Coupling}

In this section, we mainly review the torsion based gravitational
paradigm. The torsional scenario begins with a spacetime undergoing
the absolute parallelism where the parallel vector field
${h_\beta}^{a}$ determines
\begin{equation}\label{1}
{\Gamma^a}_{bc}={h_\beta}^{a}{h^\beta}_{b,c},
\end{equation}
which is a non-symmetric affine connection with vanishing curvature
and $h_{\beta a,b}=\partial_{b}h_{\beta a}$. Here we apply the Latin
indices as the notation for tangent spacetime while coordinate
spacetime indices are presented with the help of Greek letters. The
parallel vector fields (vierbein or tetrad fields) are the basic
dynamical variables which deduce an orthonormal basis for the
tangent space, i.e., $\textbf{h}_{a}\cdot\textbf{h}_{b}=\eta_{ab}$
where $\eta=\textmd{diag}(1,-1,-1,-1)$. In terms of vierbein
components, ${h^\beta}_a$, the vierbein fields are defined as
$\textbf{h}_a={h^\beta}_a \partial_\beta$ while the inverse
components ${h_a}^\beta$ meet the following conditions
\begin{equation*}
{h^a}_\beta {h_b}^{\beta}= \delta^a_b,\quad {h_a}^\beta
{h^a}_{\alpha}= \delta^\beta_\alpha.
\end{equation*}
The metric tensor is obtained through the relation
$g_{\beta\alpha}=\eta_{ab}{h^a}_\beta {h^b}_\alpha$ which gives the
metric determinant as $\sqrt{-g}=e=\textmd{det}({h^a}_\beta)$. Using
Eq.(\ref{1}), the antisymmetric part of Weitzenb\"{o}ck connection
yields
\begin{equation}\label{2}
{T^\beta}_{\alpha\sigma}={\widetilde{\Gamma}^\beta}_{~\sigma\alpha}-
{\widetilde{\Gamma}^\beta}_{~\alpha\sigma}
={h_a}^{\beta}(\partial_{\sigma}{h^a}_{\alpha}-\partial_{\alpha}{h^a}_{\sigma}),
\end{equation}
where ${T^\beta}_{\alpha\gamma}=-{T^\beta}_{\gamma\alpha}$, i.e., it
is antisymmetric in its lower indices. It is noted that the under
the parallel transportation of vierbein field, the curvature of the
Weitzenb\"{o}ck connection vanishes. Using this tensor, we obtain
contorsion tensor as
$K^{\alpha\gamma}_{~~\beta}=-\frac{1}{2}(T^{\alpha\gamma}_{~~\beta}
-T^{\gamma\alpha}_{~~\beta}-T_{\beta}^{~\alpha\gamma})$ and
superpotential tensor as
${S_\beta}^{\alpha\gamma}=\frac{1}{2}[\delta^{\alpha}_{\beta}{T^{\mu\gamma}}_{\mu}-
\delta^{\gamma}_{\beta}{T^{\mu\alpha}}_{\mu}+K^{\alpha\gamma}_{~~\beta}]$.
The torsion scalar takes the form
\begin{equation}\label{3}
T={S_\beta}^{\alpha\gamma}{T^\beta}_{\alpha\gamma}.
\end{equation}
The action of $f(T)$ gravity is given by \cite{ft,ft2}
\begin{equation}\label{4}
\mathcal{S}=\frac{1}{16\pi \mathcal{G}}\int
e\{f(T)+\mathcal{L}_m\}d^4x.
\end{equation}
The $\mathcal{G}$ is the gravitational constant and $f$ represents
the generic differentiable function of torsion scalar describing
extension of the teleparallel gravity. The term $\mathcal{L}_m$
describes the matter part of the action such as
$\mathcal{L}_m=\mathcal{L}_m(\rho,~p)$ where $\rho,~p$ are the
energy density and pressure of matter while we neglect the radiation
section for the sake of simplicity. By the variation of this action
w.r.t vierbein field, the following field equations come out
\begin{equation}\label{5}
{h_a}^{\beta}{S_\beta}^{\alpha\sigma}\partial_{\alpha}T
f_{_{TT}}+[{h_a}^{\beta}{T^\lambda}_{\alpha\beta}{S_\lambda}^{\sigma\alpha}+\frac{1}{e}
\partial_{\alpha}(h{h_a}^{\beta}{S_\beta}^{\alpha\sigma})
]f_{_T} +\frac{1}{4}{h_a}^{\sigma}f=4\pi
\mathcal{G}{h_a}^{\beta}\Theta^{\sigma}_{\beta},
\end{equation}
where the subscripts involving $T$ and $TT$ represent first and
second order derivatives of $f$ with respect to $T$ respectively.
For the sake of simplicity, we assume $8\pi \mathcal{G}=1$ in the
following.

The $f(T)$ field equations in terms of Einstein tensor gain a
remarkable importance in order to discuss various cosmological and
astrophysical scenarios \cite{ft,ft2}. This type of field equations
take place by replacing partial derivatives to covariant derivatives
along with compatibility of metric tensor, i.e.,
$\nabla_{\sigma}g_{\beta\alpha}=0$. Using the relations
$T^{\mu(\nu\gamma)}=K^{(\mu\nu)\gamma}=S^{\mu(\nu\gamma)}=0$, the
torsion, contorsion and superpotential tensor become
\begin{eqnarray*}
&&{T^\sigma}_{\beta\alpha}=
h^{\sigma}_{a}(\nabla_{\beta}h^{a}_{\alpha}-\nabla_{\alpha}h^{a}_{\beta}),\quad
{K^{\sigma}}_{\beta\alpha}=h^{\sigma}_{a}\nabla_{\alpha}h^{a}_{\beta},\\
&&{S_\sigma}^{\alpha\beta}=\eta^{ab}h^{\beta}_{a}\nabla_{\sigma}h^{\alpha}_{b}+
\delta^{\alpha}_{\sigma}\eta^{ab}h^{\tau}_{a}\nabla_{\tau}h^{\beta}_{b}
-\delta^{\beta}_{\sigma}\eta^{ab}h^{\tau}_{a}\nabla_{\tau}h^{\alpha}_{b}.
\end{eqnarray*}
The curvature tensor referred to Weitzenb\"{o}ck connection
vanishes, while the Riemann tensor related with the Levi-Civita
connection ${\Gamma^\gamma}_{\mu\nu}$ is given by
\begin{eqnarray}\nonumber
{R^\sigma}_{\beta\lambda\alpha}
=\nabla_{\alpha}{K^\sigma}_{\beta\lambda}-\nabla_{\lambda}{K^\sigma}_{\beta\alpha}+{K^\sigma}_{\tau\alpha}{K^\tau}_{\beta\lambda}
-{K^\sigma}_{\tau\lambda}{K^\tau}_{\beta\alpha}.
\end{eqnarray}
We obtain the Ricci tensor and scalar as follows
\begin{eqnarray}\label{116a}
R_{\beta\alpha}=-\nabla^{\sigma}S_{\alpha\sigma\beta}-g_{\beta\alpha}\nabla^{\sigma}
{T^\lambda}_{\sigma\lambda}-{S^{\sigma\lambda}}_{\beta}K_{\lambda\sigma\alpha},\quad
R+T=-2\nabla^{\sigma}{T^\alpha}_{\sigma\alpha},
\end{eqnarray}
where we use
${S^\alpha}_{\sigma\alpha}=-2{T^\alpha}_{\sigma\alpha}=2{K^\alpha}_{\sigma\alpha}$.
Substituting Eq.(\ref{116a}) along with Einstein tensor
$G_{\beta\alpha}=R_{\beta\alpha}-\frac{1}{2}g_{\beta\alpha}R$, we
get
\begin{equation}\label{1117a}
G_{\beta\alpha}-\frac{1}{2}g_{\beta\alpha}T=-\nabla^{\sigma}S_{\alpha\sigma\beta}
-{S^{\tau\sigma}}_{\beta} K_{\sigma\tau\alpha},
\end{equation}
Finally, inserting this equation in Eq.(\ref{5}), it yields
\begin{equation}\label{118a}
f_{T}G_{\beta\alpha}+\frac{1}{2}g_{\beta\alpha}(f-Tf_T)+Z_{\beta\alpha}f_{TT}=\Theta_{\beta\alpha},
\end{equation}
where $Z_{\beta\alpha}={S_{\alpha\beta}}^{\sigma}\nabla_{\sigma}T$.
This equation expresses a similar structure like $f(R)$ gravity at
least up to equation level and representing GR for the limit
$f(T)=T$. We take trace of the above equation, i.e.,
$Zf_{TT}-(R+2T)f_T+2f=\Theta$, with $Z={Z^\alpha}_{\alpha}$ and
$\Theta={\Theta^{\alpha}}_{\alpha}$ to simplify the field equations.
The $f(T)$ field equations can be rewritten as
\begin{equation}\label{120a}
G_{\beta\alpha}=\frac{1}{f_{_T}}\big[\Theta_{\beta\alpha}^{m}-Z_{\beta\alpha}f_{_{TT}}-\frac{1}{2}(f-Tf_{_T})\big],
\end{equation}
where $\Theta_{\beta\alpha}^{m}$ is the energy-momentum tensor
corresponding to matter Lagrangian.

Taking into account non-minimal coupling between torsion and matter,
Harko et al. \cite{21} defined the action as follows
\begin{equation}\label{act}
\mathcal{S}=\int e\{f_1+(1+\omega f_2)\mathcal{L}_m\}d^4x,
\end{equation}
where $\omega$ is the coupling constant having units of mass$^{-2}$
and $f_1,~f_2$ are arbitrary differential functions of torsion
scalar. Applying the tetrad variation on this action, we obtain the
following set of equations
\begin{eqnarray}\nonumber
&&{h_a}^{\beta}{S_\beta}^{\alpha\sigma}\partial_{\alpha}T
(f''_1+\omega
f''_2\mathcal{L}_m)+[{h_a}^{\beta}{T^\lambda}_{\alpha\beta}{S_\lambda}^{\sigma\alpha}+\frac{1}{e}
\partial_{\alpha}(h{h_a}^{\beta}{S_\beta}^{\alpha\sigma})
](f'_1+\omega f'_2\mathcal{L}_m)
\\\label{7}&&+\frac{1}{4}{h_a}^{\sigma}f_1-\frac{1}{4}\omega\partial_\alpha
T{h_a}^\beta {\mathcal{T}^{\sigma \alpha}_\beta}f'_2+\omega
f'_2{h_a}^\beta
{S_\beta}^{\sigma\alpha}\partial_\alpha\mathcal{L}_m=\frac{1}{2}(1+\omega
f_2){h_a}^{\beta}\Theta^{\sigma}_{\beta},
\end{eqnarray}
where the number of primes denotes the correspondingly order
derivative with respect to torsion scalar and ${\mathcal{T}^{\sigma
\alpha}_a}=\frac{\partial \mathcal{L}_m}{\partial\partial_\alpha
{h^a}_\sigma}$. We assume here that the matter Lagrangian
$\mathcal{L}_m$ is independent of derivatives of tetrad which
results the vanishing of ${\mathcal{T}^{\sigma \alpha}_a}$. Also,
the Bianchi identities of teleparallel gravity express the following
relationship
\begin{equation}\label{8}
\nabla_\beta\Theta_\tau^\beta=\frac{4}{1+\omega
f_2}{K^\lambda}_{\beta\tau}{S_\lambda}^{\beta\alpha}\nabla_\alpha(f'_1+\omega
f'_2\mathcal{L}_m)-\frac{\omega f'_2}{1+\omega
f_2}(\Theta^\beta_\tau-\mathcal{L}_m\delta^\beta_\tau)\nabla_\beta
T.
\end{equation}
This equation represents the substitute of energy and momentum
between torsion and matter through defined coupling. The $f(T)$
field equations for the torsion-matter coupling in the form of
Einstein tensor are given by
\begin{equation}\label{9}
G_{\beta\alpha}=\frac{1}{f_{_{1T}}+\omega
f_{_{2T}}\mathcal{L}_m}\bigg[\Theta_{\beta\alpha}^{+}-{S_{\alpha\beta}}^{\lambda}\nabla_\lambda
(f_{_{1TT}}+\omega
f_{_{2TT}}\mathcal{L}_m)-\mathcal{J}g_{_{\beta\alpha}}\bigg],
\end{equation}
where
\begin{eqnarray*}
\mathcal{J}=\frac{1}{2}\bigg\{f_1-T(f_{_{1T}}+\omega
f_{_{2T}}\mathcal{L}_m)\bigg\},\quad \Theta_{\beta\alpha}^{+}=
(1+\omega f_2)\Theta_{\beta\alpha}^m-\omega
f_{_{2T}}{S_{\alpha\beta}}^{\lambda}\nabla_\lambda \mathcal{L}_m.
\end{eqnarray*}
It is noted that the matter Lagrangian density needs to be properly
defined in the analysis of torsion-matter coupling. In the
literature for curvature-matter coupling, the proposals for this
matter Lagrangian density are as follows
\cite{25}.\\(i)~~$\mathcal{L}_m=p$ which reproduce the equation of
state for perfect fluid and proved as not
unique.\\(ii)~~$\mathcal{L}_m=-ne$ where $e$ denotes physical free
energy defined by $e=-\mathbb{T}s+\frac{\rho}{n},~\mathbb{T}$ is the
temperature, $s$ being the entropy of one particle and $n$ gives the
particle number density.
\\(iii)~~$\mathcal{L}_m=-\rho$ representing a natural
choice which gives the energy in a local rest frame for the fluid.

\section{Gravitational Field Equations for Wormhole Geometry}

In order to construct the gravitational field equations in the
framework of torsion-matter coupling, we firstly describe the
wormhole geometry. Let us assume the wormhole metric as follows
\cite{14}-\cite{20}
\begin{equation}\label{5+}
ds^2=e^{2\Psi(r)}dt^2-\frac{1}{1-\frac{b}{r}}dr^2-r^2d\phi^2-r^2\sin^2\phi
d\psi^2,
\end{equation}
where redshift function $\Psi$ and shape function $b$ are $r$
dependent functions. In order to set geometry of wormhole scenario,
some constraints are required on both of these functions. These are
describe as follows.
\begin{itemize}
\item \textbf{Shape function}: The shape of the wormhole consists of
two open mouths (two asymptomatically flat regions) in different
regions of the space connected through throat which is minimum
non-zero value of radial coordinate. This shape maintains through
the shape function $b(r)$ with increasing behavior and having ratio
$1$ with $r$. At throat, it must holds $b(r_0)=r_0$ as well as
$1-\frac{b(r)}{r}\geq0$. The flare-out condition being the
fundamental property of wormhole geometry is defined as
\begin{equation}\label{10}
\frac{1}{b^2}(b-b'r)>0.
\end{equation}
There is another constraint on the derivative of shape function at
throat, i.e., $b'(r_0)<1$ which must be fulfilled.
\item \textbf{Redshift function}: The main purpose of wormholes is to give a
way to move two-way through its tunnel which basically depends on
non-zero minimum value of $r$ at throat. For this purpose, i.e., to
keep throat open, the redshift function plays its role by
maintaining no-horizon at throat. To hold this condition, $\Psi$
must remains finite throughout the spacetime. This function
calculates the gravitational redshift of a light particle. When this
particle moves from potential well to escape to infinity, there
appears a reduction in its frequency which is called gravitational
redshift. At a particular value of $r$, its infinitely negative
value expresses an event horizon at throat. To prevent this
situation of appearing horizon so that wormhole solution may provide
traversable way, the magnitude of its redshift function must be
finite. This may be taken as $\Psi=0$ which gives
$e^{2\Psi(r)}\rightarrow 1$.
\item \textbf{Exotic matter}: The existence of wormhole solutions
requires some unusual type of matter, called exotic matter having
negative pressure to violate the null energy condition. Thus, it
becomes the basic factor for wormhole construction as the known
classical forms of matter satisfy this condition. The search for
such a source which provides the necessary violation with matter
content obeying the null energy condition occupy a vast range of
study in astrophysics.
\end{itemize}

The energy conditions originates through the Raychaudhuri equation
in the realm of general relativity for the expansion regarding
positivity of the term $R_{\beta\alpha}\mathcal{K}^\beta
\mathcal{K}^\alpha$ where $\mathcal{K}^\beta$ denotes the null
vector. This positivity guarantees a finite value of the parameter
marking points on the geodesics directed by geodesic congruences. In
terms of energy-momentum tensor, the above term under positivity
condition become $R_{\beta\alpha}\mathcal{K}^\beta
\mathcal{K}^\alpha=\Theta_{\beta\alpha}\mathcal{K}^\beta
\mathcal{K}^\alpha\geq 0$ in the framework of general relativity.
However, in modified theories of gravity, we carry effective
energy-momentum tensor which is further used in the above expression
to study energy conditions. For instance, in the underlying case,
Eq.(\ref{9}) takes the form
\begin{equation}\label{11}
G_{\beta\alpha}=\Theta^{\textmd{eff}}_{\beta \alpha},\quad
\textmd{where}\quad \Theta^{\textmd{eff}}_{\beta
\alpha}=\frac{1}{f_{_{1T}}+\omega
f_{_{2T}}\mathcal{L}_m}[\Theta_{\beta\alpha}^{+}+\Theta_{\beta\alpha}^{\textmd{TM}}],
\end{equation}
where
$\Theta_{\beta\alpha}^{\textmd{TM}}=-{S_{\alpha\beta}}^{\lambda}\nabla_\lambda
(f_{_{1TT}}+\omega
f_{_{2TT}}\mathcal{L}_m)-\mathcal{J}g_{_{\beta\alpha}}$ representing
contribution of torsion-matter coupling in the extended teleparallel
gravity. The corresponding energy condition becomes
$R_{\beta\alpha}\mathcal{K}^\beta
\mathcal{K}^\alpha=\Theta^{\textmd{eff}}_{\beta
\alpha}\mathcal{K}^\beta \mathcal{K}^\alpha\geq 0$ which is called
null energy condition. Taking into account perfect fluid with
presentation
$\Theta_{\beta\alpha}=(\rho^{\textmd{eff}}+p^{\textmd{eff}})\mathcal{U}_\beta
\mathcal{U}_\alpha-p^{\textmd{eff}}g_{\beta\alpha}$, this condition
yields $\rho^{\textmd{eff}}+p^{\textmd{eff}}\geq 0$.

At this stage, we may impose a condition on energy-momentum tensor
corresponding to matter part such that
$\Theta^{m}_{\beta\alpha}\mathcal{K}^\beta \mathcal{K}^\alpha\geq0$
to thread the wormhole while
$\Theta^{\textmd{eff}}_{\beta\alpha}\mathcal{K}^\beta
\mathcal{K}^\alpha\leq0$ gives the necessary violation. This
condition implies the positivity of energy density in all local
frames of references. Thus, it is important to study the constraints
on $\Theta^{m}_{\beta\alpha}$ in order to form wormholes. Consider
the violation of null energy condition and Eq.(\ref{11}), we get
\begin{eqnarray}\nonumber
\frac{1}{f_{_{1T}}+\omega f_{_{2T}}\mathcal{L}_m}\bigg[(1+\omega
f_2)\Theta_{\beta\alpha}^m-\omega
f_{_{2T}}{S_{\alpha\beta}}^{\lambda}\nabla_\lambda
\mathcal{L}_m-{S_{\alpha\beta}}^{\lambda}\nabla_\lambda
(f_{_{1TT}}\\\nonumber+\omega
f_{_{2TT}}\mathcal{L}_m)-\mathcal{J}g_{_{\beta\alpha}}\bigg]\mathcal{K}^\beta
\mathcal{K}^\alpha<0.
\end{eqnarray}
For the viable wormhole solutions, if $(f_{_{1T}}+\omega
f_{_{2T}}\mathcal{L}_m)>0$, then we obtain the following constraint
\begin{eqnarray}\nonumber
0\leq\Theta_{\beta\alpha}^m\mathcal{K}^\beta
\mathcal{K}^\alpha<\frac{1}{1+\omega f_2}\bigg[\omega
f_{_{2T}}{S_{\alpha\beta}}^{\lambda}\nabla_\lambda
\mathcal{L}_m+{S_{\alpha\beta}}^{\lambda}\nabla_\lambda
(f_{_{1TT}}\\\nonumber+\omega
f_{_{2TT}}\mathcal{L}_m)+\mathcal{J}g_{_{\beta\alpha}}\bigg]\mathcal{K}^\beta
\mathcal{K}^\alpha,
\end{eqnarray}
where $1+\omega f_2>0$ must holds. For the case $(f_{_{1T}}+\omega
f_{_{2T}}\mathcal{L}_m)<0$, the null energy condition
straightforwardly gives
\begin{eqnarray}\nonumber
\Theta_{\beta\alpha}^m\mathcal{K}^\beta
\mathcal{K}^\alpha>\frac{1}{1+\omega f_2}\bigg[\omega
f_{_{2T}}{S_{\alpha\beta}}^{\lambda}\nabla_\lambda
\mathcal{L}_m+{S_{\alpha\beta}}^{\lambda}\nabla_\lambda
(f_{_{1TT}}\\\nonumber+\omega
f_{_{2TT}}\mathcal{L}_m)+\mathcal{J}g_{_{\beta\alpha}}\bigg]\mathcal{K}^\beta
\mathcal{K}^\alpha.
\end{eqnarray}

We consider the case $\mathcal{L}_m=-\rho$ with no horizon
condition, that is $\Psi=0$ to construct the background geometry for
wormhole solutions in the framework of $f(T)$ gravity having
torsion-matter coupling. We take anisotropic distribution of fluid
having energy-momentum tensor as
\begin{equation}\label{13}
\Theta_{\beta\alpha}^m=(\rho+p_r)\mathcal{U}_\beta
\mathcal{U}_\alpha-p_r g_{\beta\alpha}+(p_t-p_r) \chi_\beta
\chi_\alpha,
\end{equation}
where $p_r$ is the radial directed pressure component and $p_t$
denotes tangential pressure component with $\rho=\rho(r),~p=p(r)$
satisfying $\mathcal{U}^\beta \mathcal{U}_\alpha=-\chi^\beta
\chi_\alpha=1$. Taking into account Eqs.(\ref{9}) and (\ref{5+}), we
obtain a set of field equations as follows
\begin{eqnarray}\nonumber
\frac{b'}{r^2}&=&\frac{1}{f_{_{1T}}-\omega
f_{_{2T}}\rho}\bigg[(1+\omega f_2)\rho+\frac{\omega
f_{_{2T}}\rho'}{r}\bigg(1-\frac{b}{r}-\sqrt{1-\frac{b}{r}}\bigg)\bigg]\\\label{23+}&-&\frac{T'(f_{_{1TT}}-
\omega f_{_{2TT}}\rho)}{r(f_{_{1T}}-\omega
f_{_{2T}}\rho)}\bigg(1-\frac{b}{r}-\sqrt{1-\frac{b}{r}}\bigg)-\frac{\mathcal{J}}{f_{_{1T}}-\omega
f_{_{2T}}\rho},\\\label{24+} -\frac{b}{r^3}&=&\frac{(1+\omega
f_2)p_r}{f_{_{1T}}-\omega
f_{_{2T}}\rho}+\frac{\mathcal{J}}{f_{_{1T}}-\omega
f_{_{2T}}\rho},\\\nonumber
-\frac{b'r-b}{2r^3}&=&\frac{1}{f_{_{1T}}-\omega
f_{_{2T}}\rho}\bigg[(1+\omega f_2)p_t-\frac{\omega
f_{_{2T}}\rho'}{2r}\bigg(1-\frac{b}{r}-\sqrt{1-\frac{b}{r}}\bigg)\bigg]\\\label{25+}&+&\frac{T'(f_{_{1TT}}-
\omega f_{_{2TT}}\rho)}{2r(f_{_{1T}}-\omega
f_{_{2T}}\rho)}\bigg(1-\frac{b}{r}-\sqrt{1-\frac{b}{r}}\bigg)+\frac{\mathcal{J}}{f_{_{1T}}-\omega
f_{_{2T}}\rho},
\end{eqnarray}
where prime represents derivative with respect to $r$ and
\begin{equation}\label{12}
T=\frac{2}{r^2}\bigg[2\bigg(1-\sqrt{1-\frac{b}{r}}\bigg)-\frac{b}{r}\bigg].
\end{equation}

The violation of null energy condition
$(\Theta^{\textmd{eff}}_{\beta\alpha}\mathcal{K}^\beta
\mathcal{K}^\alpha<0)$ for field equations (\ref{23+})-(\ref{25+})
is checked through the consideration of radial null vector which
yields
\begin{eqnarray}\nonumber
\rho^{\textmd{eff}}+p_r^{\textmd{eff}}&=&\frac{1}{f_{_{1T}}-\omega
f_{_{2T}}\rho}\bigg[(1+\omega
f_2)(\rho+p_r)+\frac{1}{r}\bigg\{\omega
f_{_{2T}}\rho'\\\nonumber&-&T'(f_{_{1TT}}- \omega
f_{_{2TT}}\rho)\bigg\}\bigg(1-\frac{b}{r}-\sqrt{1-\frac{b}{r}}\bigg)\bigg]\\\nonumber&=&\frac{b'r-b}{r^3}<0,\\\nonumber
\Rightarrow \rho^{\textmd{eff}}+p_r^{\textmd{eff}}&<&0,
\end{eqnarray}
where the inequality comes through the flaring out condition of
shape function. In order to discuss the above scenario at throat, we
obtain the following relationship
\begin{eqnarray}\nonumber
\rho^{\textmd{eff}}(r_0)+p_r^{\textmd{eff}}(r_0)&=&\frac{1}{f_{_{1T_0}}-\omega
f_{_{2T_0}}\rho_0}\bigg[(1+\omega
f_2(r_0))(\rho_0+p_{r0})+\frac{1}{r_0}\bigg\{\omega
f_{_{2T_0}}\rho_0'\\\nonumber&-&T'_0(f_{_{1T_0T_0}}- \omega
f_{_{2T_0T_0}}\rho_0)\bigg\}\bigg(1-\frac{b_0}{r_0}-\sqrt{1-\frac{b_0}{r_0}}\bigg)\bigg],
\end{eqnarray}
which must satisfy the following constraint in order to meet the
above inequality, that is
\begin{eqnarray*}
(1+\omega f_2(r_0))(\rho_0+p_{r0})+\frac{1}{r_0}\omega
f_{_{2T_0}}\rho_0'\bigg(1-\frac{b_0}{r_0}-\sqrt{1-\frac{b_0}{r_0}}\bigg)<&&\\\frac{T'_0}{r_0}(f_{_{1T_0T_0}}-
\omega
f_{_{2T_0T_0}}\rho_0)\bigg(1-\frac{b_0}{r_0}-\sqrt{1-\frac{b_0}{r_0}}\bigg),&&
\end{eqnarray*}
where $f_{_{1T_0}}-\omega f_{_{2T_0}}\rho_0>0$ while the inequality
reverses if $f_{_{1T_0}}-\omega f_{_{2T_0}}\rho_0<0$. In order to
write final field equations in terms of matter component, we observe
that the last term in all Eqs.(\ref{23+})-(\ref{25+}) change its
sign using signature (-,+,+,+) for wormhole spacetime while
remaining terms stay same. This leads to the vanishing of last terms
in each equation as energy density is independent of signature.
Finally, the field equations can be rewritten as
\begin{eqnarray}\nonumber
b'\big(f_{_{1T}}-\omega f_{_{2T}}\rho\big)&=&r^2(1+\omega
f_2)\rho+\omega\bigg[\omega f_{_{2T}}\rho'-T'(f_{_{1TT}}- \omega
f_{_{2TT}}\rho)\bigg]\\\label{51}&\times&\bigg(1-\frac{b}{r}-\sqrt{1-\frac{b}{r}}\bigg),\\\label{52}p_r&=&
-\frac{b}{r^3}\frac{f_{_{1T}}-\omega f_{_{2T}}\rho}{1+\omega
f_2},\\\nonumber p_t&=&\frac{1}{2r(1+\omega
f_2)}\bigg[\bigg(f_{_{1T}}-\omega
f_{_{2T}}\rho\bigg)\bigg(\frac{b}{r^2}-\frac{b'}{r}\bigg)-\bigg\{\omega\rho'
f_{_{2T}}\\\label{53}&+&T'(f_{_{1TT}}-\omega
f_{_{2TT}}\rho)\bigg\}\bigg(1-\frac{b}{r}-\sqrt{1-\frac{b}{r}}\bigg)\bigg].
\end{eqnarray}

\section{Wormhole Solutions}

In order to discuss wormhole solutions, we have to deal with
un-closed system of equations such that the three equations with
$f_1,~f_2,~\rho,~p_r,~p_t$ and $b$ are unknown functions. Due to
non-linear appearance of these equations, the explicit functions are
extremely difficult. However, we may adopt some alternative
strategies keeping in mind the characteristics of wormhole geometry.
To involve the effects of non-minimal torsion-matter coupling to
construct the wormhole solutions, we have to choose some viable
models $f_1$ and $f_2$. Now we are remaining with four unknowns for
which we may assume some kind of equation of state like equation
obeying the traceless fluid, particular value of shape function
obeying all corresponding conditions or some type of energy density
like density of static spherically symmetric smearing object, etc.
We adopt the last approach of considering energy density with
particular form \cite{25}
\begin{equation}\label{54}
\rho=\rho_0\bigg(\frac{r_0}{r}\bigg)^{\sigma},
\end{equation}
where $\rho_0$ and $\sigma$ are positive constants.

In the following, we consider two viable $f(T)$ models and study the
different conditions of shape functions as well as null energy
condition.

\subsection{Model 1}
\begin{figure} \centering
\epsfig{file=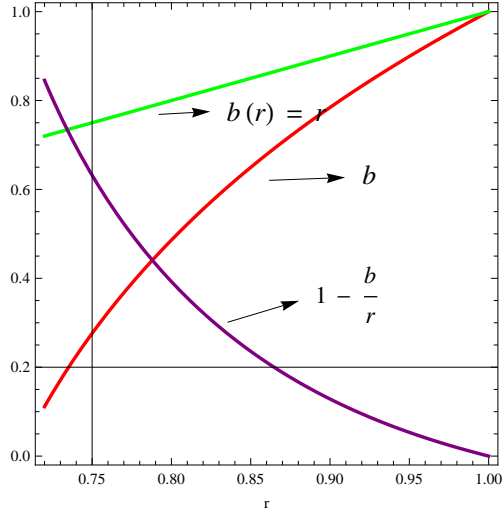,width=.50\linewidth} \caption{Plots of shape
function $b,~ 1-\frac{b}{r},~ b=r~ \textmd{versus}~ r$ for Model 1
using $r_0=1,~\rho_0=0.75,~\sigma=3$ and $\omega=0.4$.}
\end{figure}
\begin{figure} \centering
\epsfig{file=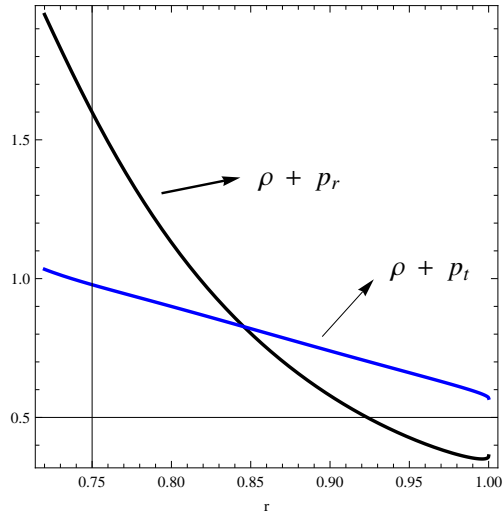,width=.50\linewidth} \caption{Plots of null
energy condition $\rho+p_r,~\rho+p_t~ \textmd{versus}~ r$ for Model
1 using $r_0=1,~\rho_0=0.75,~\sigma=3$ and $\omega=0.4$.}
\end{figure}
\begin{figure} \centering
\epsfig{file=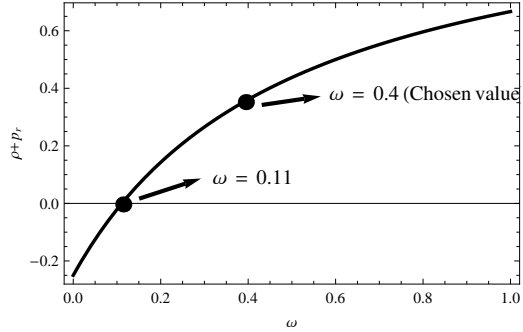,width=.50\linewidth} \caption{The general
relativistic deviation profile versus $\omega$ for Model 1 using
$r_0=1,~\rho_0=0.75$ and $\sigma=3$.}
\end{figure}
As a first model, we consider the models $f_1=T$ and $f_2=T$
analogous to the case of curvature-matter coupling scenario where
these models are taken as $f_1=R=f_2$ \cite{25}. Substituting these
values of models along with Eq.(\ref{54}) in (\ref{51}), we obtain
the following differential equation for the shape function
\begin{eqnarray}\label{55}
b'-\frac{\omega \sigma
\rho_0\big(\frac{r_0}{r}\big)^{\sigma}}{1-\omega
\rho_0\big(\frac{r_0}{r}\big)^{\sigma}}\bigg(\frac{b}{r}+\sqrt{1-\frac{b}{r}}\bigg)=-\frac{\sigma
\rho_0\big(\frac{r_0}{r}\big)^{\sigma}}{1-\omega
\rho_0\big(\frac{r_0}{r}\big)^{\sigma}}[r(1+\omega T)+\omega].
\end{eqnarray}
We plot the shape function by numerically in order to study the
wormhole geometry as shown in \textbf{Figure 1} fixing initial
condition as $b(1)=1$. We take some particular values of constants
as $r_0=1,~\rho_0=0.75,~\omega=0.4$ and plot versus $r$. The plot of
$b$ represents positively increasing behavior with respect to $r$.
The trajectory of $1-\frac{b}{r}$ depicts the positive behavior for
$r\leq1$ which meets the condition $b<r$. \textbf{Figure 2} shows
the null energy condition taking both components of pressure along
with chosen energy density. The black curve represents $\rho+p_r$
and blue curve describes $\rho+p_t$ versus $r$ taking into account
Eqs.(\ref{52})-(\ref{54}). We see that the null energy condition
holds in this case. Thus the possibility of wormhole solutions for
which the normal matter satisfying the null energy condition in the
background of torsion-matter coupling exists.

\textbf{Figure 3} represents the general relativistic deviation
profile of null energy condition for radial pressure component which
gives the range of coupling parameter. At throat, the null energy
condition implies
\begin{equation*}
(\rho+p_r)\mid_{r=r_0}=\rho_0-\frac{1-\omega\rho_0}{r_0^2+2\omega}.
\end{equation*}
For the chosen values of parameters, we obtain the range of coupling
constant as $\omega\geq0.11$ for which null energy condition holds.
This depicts that the increasing value of coupling constant
minimizes or vanishes the violation of null energy condition through
matter. For the case when there is no coupling, i.e., $\omega=0$,
Eq.(\ref{51}) reduces to $b'=-\sigma \rho_0 r_0^\sigma
r^{1-\sigma}$. The solution of this equation is $b=-\frac{\sigma
\rho_0 r_0^\sigma}{2-\sigma} r^{2-\sigma}+c$, where $c$ is an
integration constant and can be determined by $b(r_0)=r_0$. After
applying this condition, finally the shape function becomes
\begin{equation*}
b(r)=r_0\bigg(1+\frac{\sigma \rho_0
r_0}{2-\sigma}\bigg)-\frac{\sigma \rho_0 r_0^\sigma}{2-\sigma}
r^{2-\sigma}.
\end{equation*}
This shape function shows a asymptomatically flat geometry as
$\frac{b}{r}\rightarrow 0$ as $r\rightarrow \infty$. Taking same
values of parameters, this function gives $b=-1.25+\frac{2.25}{r}$
representing decreasing but positive behavior for $r<1.8$ while
$1-\frac{b}{r}$ holds for two sets of ranges $(r>-2.25~
\textmd{and}~ r>1)$ or $(r<-2.25~ \textmd{and}~ r<1)$. Also, the
condition $b'(r_0)<1$ meets as $b'=-\sigma\rho_0r_0<1$. The
expression $\rho+p_r=\frac{1}{r^4}(2r-2.25)$ leads to $r>1.125$ to
meet the null energy condition while
$\rho+p_t=\frac{1}{2r^4}(0.25r+4.5)$ which remains positive.

\subsection{Model 2}
\begin{figure} \centering
\epsfig{file=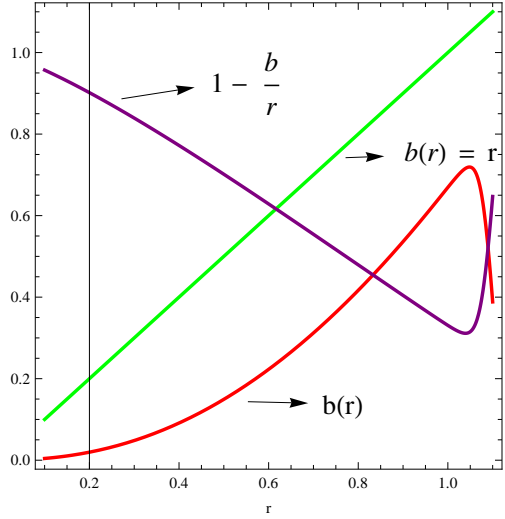,width=.50\linewidth} \caption{Plots of shape
function $b,~ 1-\frac{b}{r},~ b=r~ \textmd{versus}~ r$ for Model 2
using $r_0=1,~\rho_0=0.75,~\sigma=3,~\mu=0.2,~\nu=-2$ and
$\omega=0.4$.}
\end{figure}
As the second choice for the model, we consider the viable model
with quadratic torsion term as \cite{21}
\begin{equation}\label{m2}
f_1=-\Lambda,~\quad f_2=\mu T+\nu T^2,
\end{equation}
where $\Lambda>0,~\mu$ and $\nu$ are constants. This model describes
a well-depicted result in cosmological scenario, i.e., a
matter-dominated phase followed by phantom phase of the universe.
Substituting Eqs.(\ref{54}) and (\ref{m2}) in (\ref{51}), we obtain
the following differential equation
\begin{eqnarray}\nonumber
b'+\frac{r}{\mu+2\nu T}\bigg\{\frac{\sigma(\mu+2\nu T)}{r}-2\nu
T'\bigg\}\bigg(\frac{b}{r}&+&\sqrt{1-\frac{b}{r}}-1\bigg)=-\frac{r^2}{1+\omega(\mu+2\nu
T)}\\\label{de2}&\times&\bigg(1+\omega(\mu+\nu T)T\bigg).
\end{eqnarray}
\begin{figure} \centering
\epsfig{file=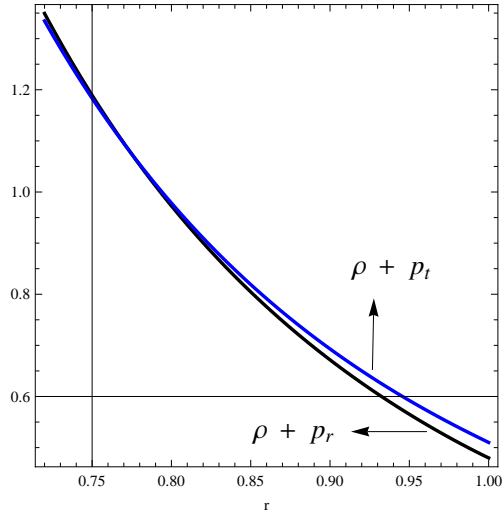,width=.50\linewidth} \caption{Plots of null
energy condition $\rho+p_r,~\rho+p_t~ \textmd{versus}~ r$ for Model
2 using $r_0=1,~\rho_0=0.75,~\sigma=3,~\mu=0.2,~\nu=-2$ and
$\omega=0.4$.}
\end{figure}
\begin{figure} \centering
\epsfig{file=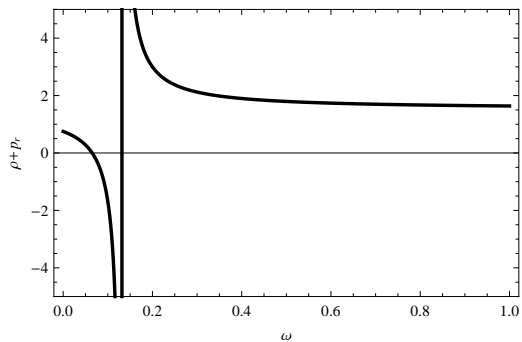,width=.50\linewidth} \caption{The general
relativistic deviation profile versus $\omega$ for Model 2 using
$r_0=1,~\rho_0=0.75,~\sigma=3,~\mu=0.2,~\nu=-2$.}
\end{figure}
Figure 4 represents the plots of shape function under different
conditions through numerical computations. The trajectory of $b$
describes positively increasing behavior for $r\geq1$ and then
decreasing behavior. The plot of $1-\frac{b}{r}$ respects positive
behavior. Thus, for model 2, we obtain wormhole geometry for
torsion-matter coupling. The null energy condition for this model is
plotted in Figure 5 which shows the positivity of the condition. In
order to check the relativistic deviation profile taking into
account null energy condition for radial pressure component at
throat, we find the equation as follows
\begin{equation*}
(\rho+p_r)\mid_{r=r_0}=\rho_0+\frac{\omega(\mu
r_0^2+4\nu)\rho_0}{r_0^4+2\omega(\mu r_0^2+2\nu)}.
\end{equation*}
Its plot is shown in Figure 6 representing $\omega>0.15$ for the
range where null energy condition holds. This depicts that the
increasing value of coupling constant minimizes or vanishes the
violation of null energy condition through matter. Also, we may
discuss the case of zero coupling in the similar way as discussed
for Model 1.

\section{Concluding Remarks}

The search for wormhole solutions satisfying energy conditions
becomes the most interesting configuration now-a-days. Wormhole is a
tube like shape or tunnel which is assumed to be a source to link
distant regions in the universe. The most amazing thing is the
two-way travel through wormhole tunnel which happened when throat
remains open. That is, to prevent the wormhole from collapse at
non-zero minimum value of radial coordinate. In order to keep the
throat open, the exotic matter is used which violate the null energy
condition and elaborated the wormhole trajectories as hypothetical
paths. In order to find some realistic sources which support the
wormhole geometry, the concentration is goes towards modified
theories of gravity. In these theories, effective scenario gives the
violation of null energy condition and matter source supports the
wormholes. In this paper, the wormhole geometries are explored
taking a non-minimal coupling between torsion and matter part in
extended teleparallel gravity. This coupling expresses the exchange
of energy and momentum between both parts torsion and matter.

The extension of $f(T)$ gravity appeared in terms of two arbitrary
functions $f_1(T)$ and $f_2(T)$ where $f_1$ is the extension of
geometric part while $f_2$ is coupled with matter Lagrangian part
through some coupling constant. At throat, the general conditions
imposed by the null energy condition taking energy-momentum tensor
of matter Lagrangian are presented in terms of non-minimal
torsion-matter coupling. The field equations appeared in non-linear
form which are difficult to solve for analytical solutions.
Presented various strategies to solve these equations, we have
adopted to assume two viable models with a non-monotonically
decreasing function of energy density. These models involving linear
torsion scalar coupled with $T$-matter and quadratic torsion term
with matter represented wormhole geometry. For these solutions, the
null energy condition is satisfied. It is concluded that that the
null energy condition is satisfied for increasing values of coupling
constant. This depicted that, the usage of exotic matter can be
reduced or vanished with the higher values of coupling constant.
Thus through the torsion-matter coupling, we have obtained some
wormhole solutions in realistic way such that matter source
satisfied the energy conditions while effective part having
torsion-matter coupled terms provided the necessary violation.
Finally, we remark here that this work may be a useful contribution
for the present theory as well as astrophysical aspects.

\end{document}